\documentclass[aps,prc,showpacs,amsmath,amssymb]{revtex4}

\usepackage{graphicx}
\usepackage{epsfig}

\begin{document}


\title{Suppression of High Transverse
  Momentum $\pi^0$ Spectra in Au+Au Collisions at RHIC.}

\author{D.~E.~Kahana}
\affiliation{31 Pembrook Dr.,
Stony Brook, NY 11790, USA}
\author{S.~H.~Kahana}
\affiliation{Physics Department, Brookhaven National Laboratory\\
   Upton, NY 11973, USA}

\date{\today}  

\begin{abstract}
Au+Au, $  {s}^{1/2} = 200$  A GeV measurements at  RHIC, obtained
with  the PHENIX,  STAR, PHOBOS  and BRAHMS  detectors,  have all
indicated   a   suppression  of   production,   relative  to   an
appropriately  normalized NN level.   For central  collisions and
vanishing  pseudo-rapidity these experiments  exhibit suppression
in  charged  meson  production,  especially at  medium  to  large
transverse  momenta.  In the  PHENIX experiment  similar behavior
has been reported for $\pi^0$ spectra.

In   a  recent   work~\cite{luc4brahms}  on   the   simpler  D+Au
interaction, to be considered perhaps  as a tune-up for Au+Au, we
reported on  a pre-hadronic  cascade mechanism which  can explain
the mixed observation of moderately reduced $p_\perp$ suppression
at higher  pseudo-rapidity as well  as the Cronin  enhancement at
mid-rapidity. Here we  present the extension of this  work to the
more massive ion-ion collisions.
  
Our major thesis is that  much of the suppression is generated in
a late stage cascade  of colourless pre-hadrons produced after an
initial short-lived  coloured phase.  We present  a pQCD argument
to  justify  this approach  and  to  estimate  the time  duration
$\tau_p$ of  this initial phase.  Of essential  importance is the
brevity in time of the  coloured phase existence relative to that
of  the   strongly  interacting  pre-hadron   phase,  the  latter
essentially an interactive  cascade.  These distinctions in phase
are  of  course  not   strict,  but  adequate  for  treating  the
suppression of moderate and high $p_\perp$ mesons.

\end{abstract}


\maketitle 
\section{Introduction}

The  specific  question  we  deal   with  in  this  work  is  the
suppression of  medium to  high transverse momentum  meson yields
observed              in             the             experimental
measurements~\cite{brahms,phenix,phobos,star}  for  Au+Au at  130
GeV and 200 GeV.  The  experiments have focused on the $\eta$ and
$p_\perp$-dependence  of the ratio  of these  yields to  those in
p+p, for charged particles:

\begin{equation}
\text{R}[\text{AA/NN}] =
\left(\frac{1}{\text{N}_{coll}}\right)
\frac{
[d^2N^{ch}/dp_\perp\,d\eta]\,\,({\text{AA}})
}
{
[d^2N^{ch}/dp_\perp\,d\eta]\,\,({\text{NN}})
}
,
\end{equation}

\noindent  where  $\text{N}_{coll}$  is  a calculated  number  of
binary NN  collisions occurring in  Au+Au at a  designated energy
and centrality. One can also,  of course, just compare the yields
directly to the data without reference to ratios.

The simulation code LUCIFER, developed for high energy heavy-ion
collisions, has previously been applied to both SPS energies
${s}^{1/2} = $(17.2,20) A GeV~\cite{lucifer1,lucifer2,luc3} and
to RHIC energies ${s}^{1/2} =$ (56,130,200) A GeV
~\cite{lucifer1,luc3}.  Although nominally intended for dealing
with soft, low $p_\perp$ interactions, it is possible to
introduce high $p_\perp$ hadron spectra via the NN inputs, which
form the building blocks of the simulations, and thus to examine
the effect of re-scattering, and concomitant energy loss, on the
high $p_\perp$ spectra~\cite{luc4brahms}.  The simulation is
divided into two phases I and II.  The first stage sets up the
participants, both mesonic and baryonic, their four momenta and
positions for the commencement of a `hadronic' cascade in phase
II. Energy loss and interactions within what amounts to a
`hadronic' fluid in phase II play a key role in the eventual
suppression of the transverse momentum distributions.

This approach has already yielded a consistent description of at
least the final stage phenomena leading to measurements in both
$D+Au$~\cite{luc4brahms} and $Au+Au$ at a variety of kinematical
conditions, energies and transverse momenta. All these results
follow given only the presence at early times during the ion-ion
collision of a fluid of `pre-hadrons' having rather generic properties.

Such a mechanism for  the suppression of high transverse momentum
jets  may seem  to run  counter  to the  conventional pQCD  view.
There does exist, however, notable justification for a picture in
which  colourless pre-hadrons,  may be  produced at  rather early
times in a AA collision, and  then play a key role in the further
development of  the system.  First  there is the work  of Shuryak
and Zahed~\cite{zahed}  on the persistence  of hadron-like states
above the  critical temperature in a dense  quark-gluon medium as
well  as  similar  results  from lattice-driven  studies  on  the
persistence   of  the   J/$\Psi$  and   other   special  hadronic
states~\cite{lattice}.  Recently a much more direct treatment has
been given, on which we  now mainly rely, by Kopeliovich, Nemchik
and  Schmidt~\cite{boris1} as  well as  Berger~\cite{berger}, who
directly consider the  temporal dynamics of pre-hadron production
from a  pQCD point  of view.  In  particular Kopeliovich  {\it et
al.},  outline the  fate of  leading hadrons  from jets,  in deep
inelastic  scattering  (DIS)  on  massive ions  and  discuss  the
relevance to relativistic heavy ion collisions.  These works have
the distinct  advantage of relying directly on  pQCD.  We attempt
to reproduce their development, as it is germane to our treatment
of NN, NA and AA interactions.

Figure (1), essentially borrowed from Reference(11), displays the
production of a jet, here at least a moderately high energy
quark, in a pp or pA event and the time scales relevant to the
process.  Kopeliovich {\it et al.}  begin by explicitly
considering eP or eA, in which case the initiating particle in
the jet creation is an off-shell photon ($\gamma^*$). In pp or AA
one could equally well employ a gluon at the first vertex.  The
kinematics~\cite{boris1} remains essentially the same then for
nuclear induced events and we elaborate this case somewhat.

The essential feature evident in the figure is the production of
a colour neutral pre-hadron after a time $t_p$, from an initially
perturbative high $p_\perp$ quark, by what one could label as
coalescence of the leading quark with an ambient anti-quark.
This coalescence incidentally is most probable for co-moving
anti-quarks, the scarcity of co-movers with increasing $p_\perp$
explaining the rapid fall off of even the pp meson spectrum with
increasing $p_\perp$.  Also prominent in this diagram are the
early perturbative gluon radiation.  These radiated gluons may in
due course create further pre-hadrons, and of course there is
then a concomitant energy loss from the initial quark.  Further,
for leading partons, one expects that the hadronisation time
$t_h$ is considerably longer than $t_p$, involving soft processes
that eventually put the pre-hadron on its final hadronic mass
shell.  Octet $q \bar q$'s might also arise but will not persist
or eventually coalesce given the repulsive or non-confining
forces in such  systems.

One can summarise the kinematic arguments~\cite{boris1} most
easily in the rest frame of the struck proton or nucleus. We use
these arguments to estimate the production time $t_p$ for the
pre-hadron, which will be a considerably shorter time $\tau_p$ in
the colliding frame of an AA system. For DIS on a nucleon, on its
own or within a nucleus A, the production time in Figure (1) is
estimated to be

\begin{equation}
t_p \sim \frac{[E_q]}{[dE_q/dz]}(1-z_h),
\end{equation}
\noindent
where  $z_h$ is  the fraction  of  quark energy  imparted to  the
hadron. Integrating the gluon  radiation spectrum one obtains for
the  energy  loss  per   unit  length  $z$,  a  time  independent
rate~\cite{boris1,niedermayer} rising quadratically with the hard
scale $Q$,

\begin{equation}
-\frac{DE}{Dz} = \frac{2\alpha_s}{3\pi}Q^2.
\end{equation}

The colour neutralisation time is then given by:

\begin{equation}
t_p \sim
\frac{E_q}{Q^2} (1-z_h).
\end{equation}
\noindent
The hadronisation time is related to the QCD scale
factor and is usually estimated as:
\begin{equation}
t_h \sim
\frac{E_h}{\Lambda^2_{QCD}} (1-z_h),
\end{equation}

\noindent and is much longer than the colour neutralisation time,
given that the QCD scale is close to 200 GeV.

The  authors   of  Reference  (11)  argue  the   brevity  of  the
de-colourisation   time  scale   is  a   consequence   of  energy
conservation, coherence and Sudakov suppression.

Importantly,  a much  reduced  scale for  pre-hadron creation  is
likely  to  remain true  even  for  non-leading partons  provided
Equation (4) remains valid, {\it i.~e.} when $Q$ is comparable to
$E_q$.  Thus  many of the  radiated hard or even  moderately hard
gluons, in the early stages of a pp or AA collision will initiate
similar  pre-hadrons, and in  a nuclear  medium such  large sized
objects will be sufficiently  numerous and of critical importance
to  the dynamics.   One  should  keep in  mind  that the  overall
multiplicity is not large at  $p_\perp \ge$ 1 GeV/c, where the NN
spectra (see  Figure (5))  has dropped from  that at  the softest
transverse momenta  measured by close to two  orders of magnitude
and for Au+Au (Figure 7) by somewhat more.

One concludes that the production and hadronisation processes are
to some degree separate: with generally $t_p$ less than $t_f$ and
frequently much less. Corresponding time scales in, say, the
center of momentum frame for an A+A system will of course be
considerably contracted. Although the colourless pre-hadron in
Figure (1) is generated by an initially perturbative process with
an anti-quark, it is the subsequent interactions with other such
pre-hadrons that leads to the observed suppression for mesons of
appreciable and even moderate transverse momentum.  The
pre-hadron perturbatively begins life with the $q\bar q$ at small
relative distance, and thus has a small initial size, but in
light of the ambient momenta for such high or even moderately
high energy partons, the transverse diameter rapidly increases to
the scale of a typical hadron. Reference (11) suggests the entire
growth to pre-hadronic size occurs quickly.

One can pursue the evolution of the system of pre-hadrons via a
Glauber-theory~\cite{boris1,glauber} based treatment of their
interactions, or, as we do, via a standard cascade model.  The
end result is little different: what has been labelled jet
suppression results.  In Glauber theory the hadron-sized
cross-sections produce strong absorption: and hard pre-hadrons
simply do not remain in the final state, they are too often
absorbed.  In the cascade described hereafter the pre-hadronic
medium is sharply cooled by the interactions and instead of on
hard meson, many soft mesons appear, and at lower $p_\perp$.

To obtain the final cascade yields for both
D+Au~\cite{luc4brahms} and Au+Au, it is essential that the
characteristic time $t_p$ be considerably less than $t_f$.
Indeed this constraint plus the appropriately large pre-hadron
interaction cross-sections are also instrumental in generating
the surprisingly large elliptical flow that has been measured at
RHIC~\cite{molnar,flow}.  References (11,12) provide a
perturbative QCD justification for both such conditions, the
early de-colourisation and the large pre-hadron cross-sections.
    
For completeness we present a brief overview of the dynamics of
our Monte Carlo simulation, which has already been described
extensively in earlier works~\cite{luc4brahms,lucifer2}.  Many
other simulations of heavy ion collisions exist and these are
frequently hybrid in nature, using say string models in the
initial
state~\cite{rqmd,rqmd2,bass1,frithjof,capella,werner,ko,ranft}
together with final state hadronic collisions, while some are
either purely or partly
partonic~\cite{boal,eskola,wang,wang2,geiger,bass2} in nature.
Our approach is closest in spirit to that of RQMD and
K.~Gallmeister, C.~Greiner, and Z.~Xu~\cite{greiner} as well as
work by W.~Cassing~\cite{rqmd,greiner,cassing} and
G.~Wolschin~\cite{wolschin}.  Certainly our results seem to
parallel those of the latter authors.  They seek to separate
initial perhaps parton dominated processes from hadronic
interactions occurring at some intermediate but not necessarily
late time, or to rely on some hadronic dynamics.

The ostensible purpose of describing such high energy collisions
without explicitly introducing the parton substructure of
hadrons, at least for soft processes, was to set a baseline for
judging whether deviations existed between simulation and
experiment, which could then signal interesting phenomena.  The
division between soft and hard processes is not necessarily easy
to identify in heavy ion data, although many authors believe they
have accomplished this within a gluon-saturation
picture~\cite{saturation,cgc1,cgc2}.  For both D+Au and Au+Au
systems we separate the effects of a second stage cascade, a
lower energy but still early hadronic cascade, from those of the
first stage, a parallel rather than sequential treatment of
initial (target)-(projectile) NN interactions.  The first, quite
likely partonic, stage ends after some average time $t_p$.
Paradoxically for us then, it is perhaps higher $p_\perp$ meson
production that is most justified by pQCD.

In the present work we do indeed find considerable suppression of
the Au+Au transverse momentum  spectrum at central rapidity.  The
key dynamics  occur presumably after the  de-colourisation phase in
Reference  (11), i.~e.  after  some time  closely related  to the
much  contracted  production  time  in the  collision  center  of
momentum frame $\tau_p \sim \gamma_{cm}^{-1}t_p$.

The first stage of our simulation, involving the collective
interaction of the initially present nucleons, produces a `hot'
gas or liquid of pre-hadrons which is considerably cooled in an
inevitably expanding and interacting final state cascade.  This
cooling produces the observed `jet' suppression, and not
surprisingly, that suppression is appreciably greater for Au+Au
than for D+Au~\cite{brahms,luc4brahms}.  It is clear that the
early pre-hadron `medium' resembles a rather dense fluid.

\section{The Simulation}
\subsection{Phase I}

Phase I of LUCIFER considers the initial interactions between the
separate nucleons in the colliding ions A+B, but is not a
cascade. It amounts to a creation of the initial conditions for
the final lower energy cascade.  The totality of events in phase
I involving each projectile particle happen essentially
simultaneously or one might say in parallel.  Energy loss due to
pre-hadron production as well as the creation of transverse
momentum ($p_\perp$), initial conditions for what follows, are
occur in phase I.  A model of NN
collisions~\cite{lucifer1,lucifer2}, incorporating most known
inclusive cross-section and multiplicity data, guides phase I.
The two body model, clearly an input to our simulation, is fitted
to the elastic, single-diffractive (SD) and
non-single-diffractive (NSD) aspects~\cite{goulianos} of high
energy $pp$ collisions ~\cite{ua5,ua1} and $P\bar P$
data~\cite{fermilab}.  The soft QCD aspects of early A+B events
are then modeled through their incorporation in the
phenomenological nucleon-nucleon inputs.

One notes it is precisely the energy dependence of the
cross-sections and multiplicities of the NN input that led to our
successful {\it prediction}~\cite{lucifer2} of the rather small
($\sim 13\%$) increase in $dN^{ch}/d\eta$ that was first observed
between ${s}^{1/2} = 130$ and $s^{1/2}200$ A GeV in the PHOBOS
data~\cite{phobos1}.

A history of the collisions that occur between nucleons as they
move along straight lines in phase I is recorded and later used
to guide the determination of multiplicity.  Collision driven
random walk fixes the $p_\perp$ to be ascribed to the baryons at
the start of phase II.  The overall multiplicity, however, is
subject to a modification, based as we believe on natural
physical requirements~\cite{lucifer1}.

The collective/parallel method of treating many NN collisions
between the target and projectile is achieved by defining a group
structure for interacting baryons, which preserves the geometry
of the nucleus-nucleus collision.  This is best illustrated by
considering a prototypical proton-nucleus (pA) collision.  A
group is defined by spatial contiguity.  A nucleon at some impact
parameter $b(\bar{x}_\perp)$ is imagined to collide with a
corresponding `row' of nucleons sufficiently close in the
transverse direction to the straight line path of the proton,
{\it i.~e.}~within a distance corresponding to the NN
cross-section.  In a nucleus-nucleus (AB) collision this
procedure is generalized by making two passes: on the first pass
one includes all nucleons from the target which come within the
given transverse distance of some initial projectile nucleon,
then on the second pass one includes for each target nucleon so
chosen, all of those nucleons from the projectile approaching it
within the same transverse distance.  This totality of mutually
colliding nucleons, at similar transverse displacements,
constitute a group.  The procedure partitions target and
projectile nucleons into a set of disjoint interacting groups as
well as a set of non-interacting spectators in a manner depending
on the overall geometry of the AB collision.  Clearly the
largest groups in pA will, in this way, be formed for small
impact parameters $b$; while for the most peripheral collisions
the groups will almost always consist of only one colliding pN
pair. Similar conclusions hold in the case of AB collisions.

In phase II of the cascade we treat the re-scattering of the
pre-hadrons produced in phase I.  These pre-hadrons, both
baryonic and mesonic in type, are not physical hadron resonances
or stable particles such as appear in the particle data tables,
which can only materialize after the hadronisation time $t_h$ is
passed.  Importantly, pre-hadrons, are allowed to interact
starting at earlier times, {\it viz.}  $\tau_p$ is comparable in
fact to the nominal target-projectile crossing time $T_{AB} \sim
2R_{AB}/\gamma$.  In practice we vary the effective time delay
for pre-hadron collisions in phase II, between 0.2 and 0.4 fm/c
in the center of velocity frame.

The pre-hadrons, when mesonic, may consist of a spatially close,
loosely correlated quark and anti-quark pair, and are given a
mass spectrum between $m_\pi$ and $1.2$ GeV, with appropriately
higher upper and lower mass ranges for pre-hadrons including
strange or heavier quarks.  The Monte-Carlo selection of masses
is then governed by a Gaussian distribution,
\begin{equation}
P(m)= \exp(-(m-m_0)^2/w^2),
\end{equation}
with $m_0$ a selected  center for the pre-hadron mass distribution
and $w=m_0/4$ the width.  For non-strange mesonic pre-hadrons $m_0
\sim 800$ MeV,  and for strange $m_0\sim 950$  MeV. Small changes
in $m_0$ and $w$ have little effect since the code is constrained
to fit hadron-hadron data using Equation (6).

Too high an upper limit for $m_0$ would destroy the soft nature
expected for most pre-hadron interactions when they finally decay
into `stable' mesons.  The mesonic pre-hadrons have isospin
structure corresponding to $\rho$, $\omega$, or $K^*$, while the
baryons, in present calculation, are constrained to range across
the octet and decuplet.

\subsection{Elementary Hadron-Hadron Model}

The underlying NN interaction structure involved in phase I has
been introduced in a fashion dictated by standard proton-proton
modeling~\cite{goulianos}.  The components of this modeling,
elastic, single diffractive (SD) and non-single diffractive (NSD)
were introduced above.  Fits are obtained to the existing
two-nucleon data over a broad range of energies $s^{1/2}$, using
the above pre-hadrons as an intermediate stage. No re-scattering,
only decay of these intermediate structures is permitted in the
purely NN calculation.  Cross-sections in pre-hadron collisions
were assumed to be the same size as hadron cross-sections, {\it
\it i.e.} the same as the meson-baryon or meson-meson
cross-section, at the same center of mass energy.  Where
hadron-hadron cross-sections are inadequately known we employ
straightforward quark counting to estimate the
scale. Specifically, meson-meson interactions are scaled to $4/9$
of the known NN cross-sections, thus no new parameters are
invoked.  Indeed, since known data then constrains the
pre-hadronic interaction, this approach is a parameter-free input
to the AA dynamics.

The explicit fit to these cross-sections is displayed in Figure
(2). To achieve this representation we employed, for example, a
parametrization of the data at high energies which actually works
down to 5 GeV:

\begin{equation}
\sigma_{tot} =
35.43-\frac{33.4}{s^{0.545}} +\frac{45.23}{s^{0.458}} +
0.308\,(-3.364+\ln(s))^2, 
\end{equation}
taken from V.~Ezhela {\it et~al.}~\cite{COMPAS}. 

We have  also created an extended  fit to the  elastic data using
higher   energy   CDF   measurements   at  546   GeV   and   1800
GeV~\cite{CDF}:
\begin{equation}
\sigma_{el} =
7.0 - \frac{10.0}{s^{0.545}} + \frac{10.0}{s^{0.458}} +
0.09\,(-3.364+\ln(s))^2. 
\end{equation}

The second important aspect of the NN interactions are the
multiplicities of produced particles, which are again taken from
measured and analyzed data.  We employ KNO~\cite{kno} scaling,
with very little loss of accuracy for the transverse momenta of
leading mesons, presumably from jets; {\it i.~e.} the probability
of achieving a given multiplicity at a certain energy is simply a
function of the ratio of that multiplicity to the mean
multiplicity.  To illustrate the fits, Figures (3,4) show two
energies, bracketing those relevant to RHIC physics.  Equally
good representations of intervening energy multiplicity data are
of course obtained.  As noted by many
authors~\cite{intermittency} the KNO scaling begins to break down
for the tails of the highest energy distribution (see Fig.3), but
in a fashion which again would have very little quantitative
effect on our presentation.

\subsection{High Transverse Momenta}   
           
A question that has yet to be addressed concerns the high
$p_\perp$ tails included in our fits to NN interactions and thus
in our calculations.  In fact it is certainly to such high
$p_\perp$ produced particles that the arguments of Reference (11)
best apply.  We simply include high $p_\perp$ meson events
through inclusion in the basic hadron-hadron interaction which is
of course an input to, rather than a result of, our simulation.
Thus in Figure(5) we display the NSD $(1/2\pi
p_\perp)(d^2N^{charged}/dp_\perp\,d\eta)$ from UA1~\cite{ua1}.

One can use a single exponential together with a power-law tail
in $p_\perp$, or alternatively two exponentials, to achieve a fit
to UA1 ${s}^{1/2}$=200 GeV data. A sampling function of the form
\begin{equation}
f = p_\perp (a \exp(-p_\perp/w) + b / ((1 + (r / \alpha)^ \beta)),    
\end{equation}
gives a  satisfactory fit to the pp data in the Monte-Carlo.

Additionally, since we, for the moment, constrain our comparisons
to the production of neutral pions in Au+Au we also present, in
Figure (6), the PHENIX~\cite{phenixpp} mid-rapidity $p_\perp$
yield for pp together with our representation of this spectrum.
 
The previous theoretical discussion~\cite{boris1,berger} strongly
suggests that even moderately high $p_\perp$ pre-hadrons, perhaps
as low as $1$ GeV/c, should be given hadron-like cross-sections.
Furthermore, the earliest meson-meson collisions in our second
phase cascade have an initial peak rate (in CM time) at $\sim
.25$--$.35$ fm/c, with many collisions occurring at considerably
later times.  This permits even smaller pre-hadrons to
appreciably increase their transverse size before colliding and
simultaneously suggests that most collisions occur between
co-moving pre-mesons.

\subsection{Initial Conditions for Phase II}                  

The final operation in phase I is to set the initial conditions
for the cascade of pre-hadrons in phase II. Again we are guided
in this by the elementary NN data mentioned above.  The
energy-momentum taken from the initial baryons and shared among
the produced pre-hadrons is established and an upper limit is
placed on the multiplicity of pre-hadrons.  A final accounting of
energy sharing is imposed through an overall 4-momentum
conservation requirement.

The spatial positioning of the particles at this time could be
accomplished in a variety of ways.  For an AB collision we chose
to place the pre-hadrons from each group inside a cylinder with
transverse dimension determined by the group itself, given the
initial longitudinal size of the interaction region at each
impact parameter.  We then allow the cylinder to evolve freely
according to the longitudinal momentum distributions, for a fixed
time $\tau_p$, defined in the rest frame of each group.  At the
end of $\tau_p$ the total multiplicity of the pre-hadrons is
limited so that, if given normal hadronic sizes appropriate to
meson-meson cross-sections $\sim (2/3)(4\pi/3) (0.6)^3$ fm$^{3}$,
the pre-hadrons {\it do not overlap} within the cylinder.  Such a
limitation in density is consonant with the general notion that
produced hadrons can only exist when separated from the
interaction region in which they are generated~\cite{gottfried}.
One can conclude from this that the pre-hadron matter acts like
an incompressible fluid, {\it viz.}  a liquid, a state described in the
earliest calculations with LUCIFER~\cite{lucifer1,lucifer2}.

Up to this point longitudinal boost invariance is completely
preserved since phase I is carried out using straight line paths.
The technique of defining the evolution time in the group rest
frame is essential to minimizing residual frame dependence which
inevitably arises in any cascade, hadronic or partonic, when
transverse momentum is present.  This dependence is due to the
finite size of the colliding objects, implied by their non-zero
interaction cross-section.

The collision history enumerates the number of interactions
suffered by each baryon and allows us to assign baryon transverse
momenta through random walk.  Pre-hadron multiplicities, subject
to the density restriction described above, are obtained using
the KNO scaling we invoke for the elementary NN interactions and
the known dependence on flavour. Production of baryon-antibaryon
pairs is allowed, guided again by constraints from NN data.
Transverse momentum is generated for the pre-hadrons and other
produced particles, paying attention to the random walk increase
for mesons created by multiple baryon-baryon collisions.
Finally, overall energy-momentum conservation is imposed and with
it the multi-particle phase space defined.

\section{Phase II: Final State Cascade}

We note that phase II is a relatively lower energy cascade with
the collisions between pre-hadrons, for example, involving
energies less than $s^{1/2}\sim 15$ GeV and even lower average
energy.  The sharing of energy-momentum in the original baryon
groups with the produced mesonic pre-hadrons leads to appreciable
energy loss from the initial baryons. This loss mirrors the
initial gluon radiation energy losses described in the pQCD
approach and as has been made clear occurs mostly before the
characteristic time $\tau_p$.  Phase II is as stated a standard
sequential cascade calculation in which the pre-hadrons interact
and decay as do any normal hadrons either already present or
produced during this cascade.  This cascade of course imposes
exact energy-momentum conservation at the level of two body
collisions.  For Au+Au, the effect of the pre-hadron interactions
is truly appreciable, greatly increasing multiplicities and total
transverse energy, $E_\perp$ both through production and via
eventual decay into the stable meson species, but also, as will
be seen, leading to the suppression of high $p_\perp$ hadron
yields.

We are then in a position to present results for Au+Au collisions
at $200$ A-GeV.  These appear in Figures (7--10) for various
double differential transverse momenta spectra and their ratios.
Most contain comparisons with PHENIX~\cite{phenix,phenixpp}
$\pi^0$ measurements.  In future work we will consider also
charged particle data, where produced proton spectra would play a
larger role with increasing $p_\perp$. One expects however to see
similar global behaviour for charged mesons.

\section{Results: Comparison with $\pi^0$ Data}

Figures (5) and (6) show the input, nucleon-nucleon data for
elementary production of $\pi^0$'s and charged particles. These
transverse momentum spectra at mid-rapidity have been compared to
results from PHENIX~\cite{phenixpp} and UA1~\cite{ua1}.  Figure
(7) contains the simulated $\pi^0$ transverse momenta spectrum
for Au+Au at $\eta=0$ alongside the PHENIX data~\cite{phenix}.
To a large extent the suppression observed experimentally is
paralleled by the simulated calculations. The production time
$\tau_p$ introduced above was given two values
$(2R_{Au}/\langle\gamma\rangle)$ and twice this value.  The
variation with this initial state time, a parameter in our
modeling, was small.  $\langle\gamma\rangle$'s here is the
longitudinal Lorentz factor defined above and introduced for each
baryon group separately in its rest frame.  

It's evident from the figures that, most emphatically, one {\sl
cannot} ignore final state cascading.  Moreover, with the
assumptions we have made, the most crucial being perhaps the
early commencement time for such cascading, the suppression
cannot be considered as necessarily a sign for production of a
quark-gluon plasma: the results may perhaps equally well signify
only the creation of a pre-hadron dominated medium after a short
initial delay $\sim \tau_p$.  We repeat: despite the apparently
short time $\tau_p$ at say $s^{1/2}$ =200 GeV, one finds in
practice that the phase II collisions have an effective
production or delay time of $0.25-0.35$ fm/c and continue for
some tens of fm/c. `Co-moving' collisions dominate the cascade.

It's instructive to dissect the contributions of the different
phases of the simulation, {\it i.e.} to separately show the
spectrum at the conclusion of phase I from that which results
after both phases I and II are complete.  In Figure (8) the
$\pi^0$ transverse momentum yield is shown for both these cases
against the experimental data.  It is immediately evident that
the many virtual NN collisions in phase I produce a much elevated
$p_\perp$ output and that this is in turn reduced by more than an
order of magnitude by collisions with other pre-hadrons in phase
II. Part of this effect is through inclusion in the dynamics of
at least a kinematical treatment of energy loss.  Thus the above
analogy of an initial hot gas cooled by final state interactions
during expansion, is apt.

One might well turn this around and declare that the final state
scattering of a given pre-hadron with its co-movers has cut down
the Cronin effect, a reduction which suggests the applicability
of the term `jet suppression' through final state
interactions. One notes parenthetically that particles lost at
high $p_\perp$ are compensated for by an increase in those
appearing at the lowest $p_\perp$'s. This is of course part of
the effective cooling observed.

A second and equally important criterion for the simulation is
the maximum density imposed on initial conditions for phase II.
Figure (9) casts some light on this and on another issue, the
actual transverse energy density attributed to the earliest
stages of the collision.  We have included in this figure the
charged $dN/d\eta$ spectra, including a BRAHMS charged meson
result (compared to a simple fit):

\begin{list}{}{}
\item[(a)] {for the totality of `stable' mesons in phase I+II}

\item[(b)] {the same result for phase I alone when only decays of
pre-hadrons but no phase II re-interactions are permitted,}

\item[(c)] pre-hadrons in phase I with no decays
\end{list}

It is evident that some $2/3$ of the summed tranverse energy
$E_\perp$ is generated in the second expanding phase II when the
system is increasing both longitudinally and transversely.  The
initial, very early, $E_\perp$ generated is then reduced
commensurately, falls well below the Bjorken limit and is hence
not all available for initial coloured `plasma' generation.  In
present calculations at the initialisation of II, and keeping in
mind the average masses assigned to pre-hadrons centered at $0.8$
to $1.0$ GeV, the ambient transverse energy densities are $\le
1.8$ GeV/$fm^3$) for the shortest initial time $\tau_p$ chosen
and correspondingly less for longer times.  Lattice
calculations~\cite{lattice}, however, do consider such densities
adequate for observation of the so-called cross-over transition.

\subsection{Ratios of Au+Au to NN Production}

In Figure (10) we display the often discussed
ratio~\cite{phenix4,star, brahms,phobos} introduced in Eqn.(1).  The
discrepancies between simulation and data are magnified but the
general agreement shown in Figure (3) remains intact.  The direct
double-differential cross-sections fall many orders of magnitude over
the measured range of $p_\perp$ but the ratios to normalized
proton-proton data cover a much compressed range of $p_\perp$.  In
this figure we presented PHENIX $\pi^0$ data from an early
paper~\cite{phenix}, but more recent data~\cite{shimamura} tells a
somewhat modified tale. The dip in R[AuAu/NN] at the lowest transverse
momentum point $p_\perp$= 1.25 GeV/c, implying a maximum in this ratio
at slightly higher $p_\perp$, has apparently disappeared.  This is so
for both the $0-5\%$ and $0-10\%$ central data resulting in curves in
better agreement with the corresponding theoretical calculations at
the measured $p_/perp$'s.  The equivalent charged particle
ratios~\cite{brahms,phenix4,phobos,star} still exhibit the apparent
maximum, but in a region where the proton spectra show considerable
activity. One might also question the use of the ratio R[AA/NN] at
very low $p_\perp$, involving as it does the number of binary
collisions.  A relevant or preferred divisor there might be the number
of participants.

The pQCD arguments given above for the early de-colourisation of the
initial medium would only apply to sufficiently hard mesons, perhaps
$p_\perp \sim 1$ GeV/c or greater. Nevertheless, our soft hadronic
approach for the lower transverse momenta does apparently give a good
description of the spectra at the lowest $p_\perp$'s as well.

In Figure (10) we present LUCIFER calculations for $0\%-10\%$
centrality, but the shape is generic at least for reasonably central
collisions.  As one approaches extreme peripherality the ratio will at
all $p_\perp$ eventually approach unity, provided of course the number
of binary collisions $N_{coll}$ is also adjusted.

It's worth repeating that no additional adjustable parameters, beyond
the times for production, $\tau_p$, and hadronisation (decay of
pre-hadrons), $\tau_f$, were, nor would such an approach be consistent
with our goals.  We seek only a qualitative and reasonably
quantitative understanding of the observations.

The simulations performed here overestimate, but not by very much, the
suppression of the spectrum at the lowest $p_\perp$, but yield an
overall, perhaps surprisingly, accurate description.

\section{Conclusions}

An alternative interpretation of the high transverse momentum
suppression in Au+Au has been presented. We have employed a pQCD
argument to justify the production of pre-hadrons at early times
within our simulation, and the presence of relatively large objects
leads to a great deal of scattering in the final state cascade.  It's
this rescattering that produces the `jet' suppression in our picture.
Of course, and paradoxically, it is to the higher $p_\perp$ mesons at
RHIC that the pQCD argument best applies. It is, however, interesting
to extend our treatment to the lower reaches of the spectrum for the
ratio $R[AuAu/NN]$.  Certainly there is still a silent elephant
lurking in the background: the observation of rather large elliptical
flow in the meson spectrum~\cite{flow}.  These flow measurements are
most easily reproduced theoretically in parton cascade models when
hadron-sized cross-sections are invoked~\cite{molnar}, and
cross-sections of this magnitude are what our pre-hadron cascade
employs.

In the work on D+Au~\cite{luc4brahms} the use of such a pre-hadron
spectrum exposed most clearly the simple role of dynamically-driven
geometry in ratios of BRAHMS $p_\perp$ spectra at varying
$\eta$, suppression in this case actually occurring only at
high $\eta$.

For Au+Au the modification of the $p_\perp$ spectra during phase II of
LUCIFER is considerably increased and suppression is seen even at
$\eta=0$.  Certainly, the RHIC experiments are creating unusual
nuclear matter, at high hadronic and energy density, with exciting
consequences. Surely more experimental exploration is required.

One important, if controversial, aspect of our combined pQCD
justification of pre-hadrons together with their subsequent phase II
cascade, is the likelihood that hadronic measurements are not
necessarily probing the earliest coloured stages of the AA
interaction.  To illuminate this coloured phase it may be necessary to
invoke dilepton or direct $\gamma$ signals.  In fact $10$ GeV jets
are, for the full RHIC energy ${s}^{1/2} = 200$ GeV, not perhaps the
`purest' jets in a pQCD approach, and one can perhaps reach even
higher transverse momentum, $p_\perp > 10-12$ GeV/c, for signs of
deviation from the so far observed suppression.

\section{Acknowledgements}
This manuscript has been authored under the US DOE grant
NO. DE-AC02-98CH10886. One of the authors (SHK) is also grateful to
the Alexander von Humboldt Foundation, Bonn, Germany and the
Max-Planck Institute for Nuclear Physics, Heidelberg for continued
support and hospitality.  Useful discussion with the BRAHMS, PHENIX,
PHOBOS and STAR collaborations are gratefully acknowledged, especially
with C.~Chasman, R.~Debbe, F.~Videbaek.  D.~Morrison,
M.~T.~Tannenbaum, T.~Ulrich and J.~Dunlop.

\vfill\eject

\begin{figure}
\vbox{\hbox to\hsize{\hfil
\epsfxsize=6.1truein\epsffile[0 0 561 751]{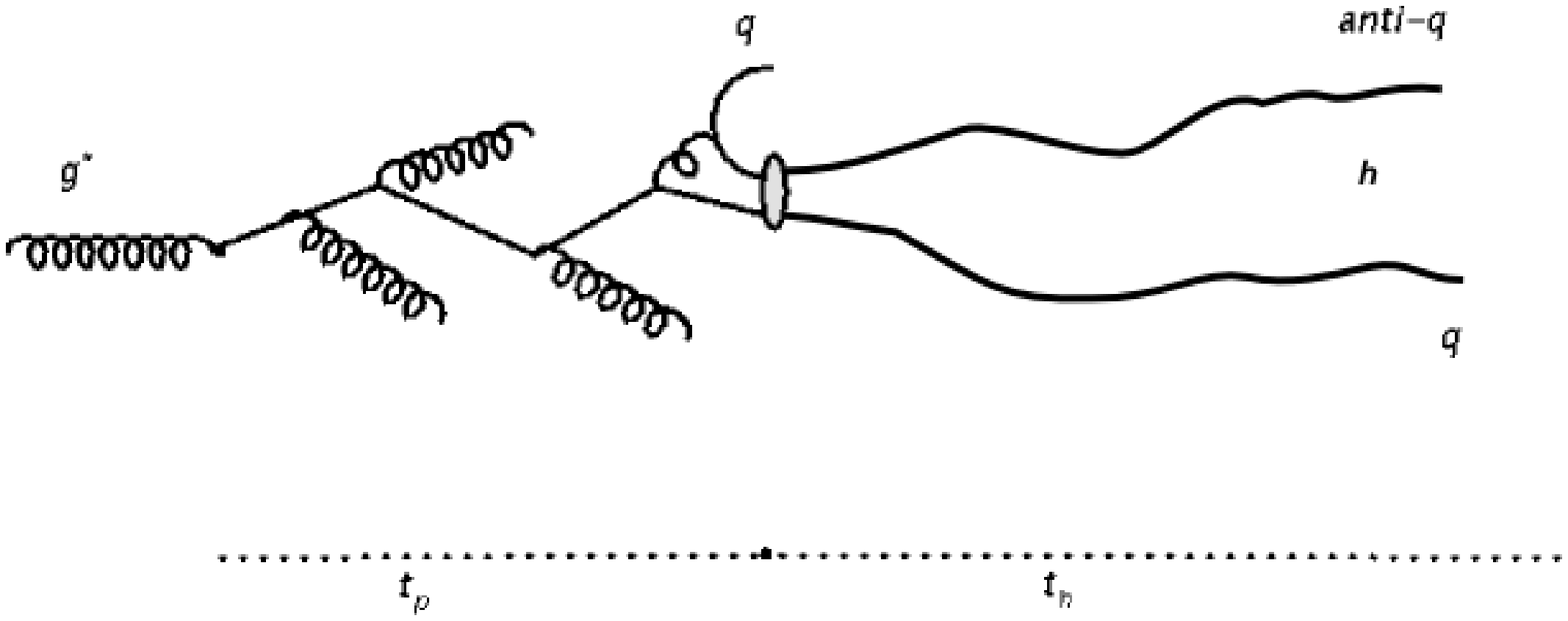}
\hfil}}
\caption[]{Schematic drawing, borrowed from Kopeliovich et.
al~\cite{boris1}, of the perturbative formation of a pre-hadron from
an off-shell gluon $g^*$ incident on a quark in the rest frame of
nucleus B in an A+B or P+B event.  The quark radiates gluons and
eventually combines with a perturbatively produced anti-quark to form
at first a small colourless pre-hadron which rapidly expands to
hadronic size.  The time $t_p$ signifies the production time of the
pre-hadron and $t_h$ its later hadronisation time. It is argued in the
text that in general $t_p$ is much less than $t_h$, and the existence
of two such time scales is critical to the parton--hadron dynamics.}
\label{fig:Fig.(1)}
\end{figure}
\clearpage

\begin{figure}
\vbox{\hbox to\hsize{\hfil
\epsfxsize=6.1truein\epsffile[0 0 561 751]{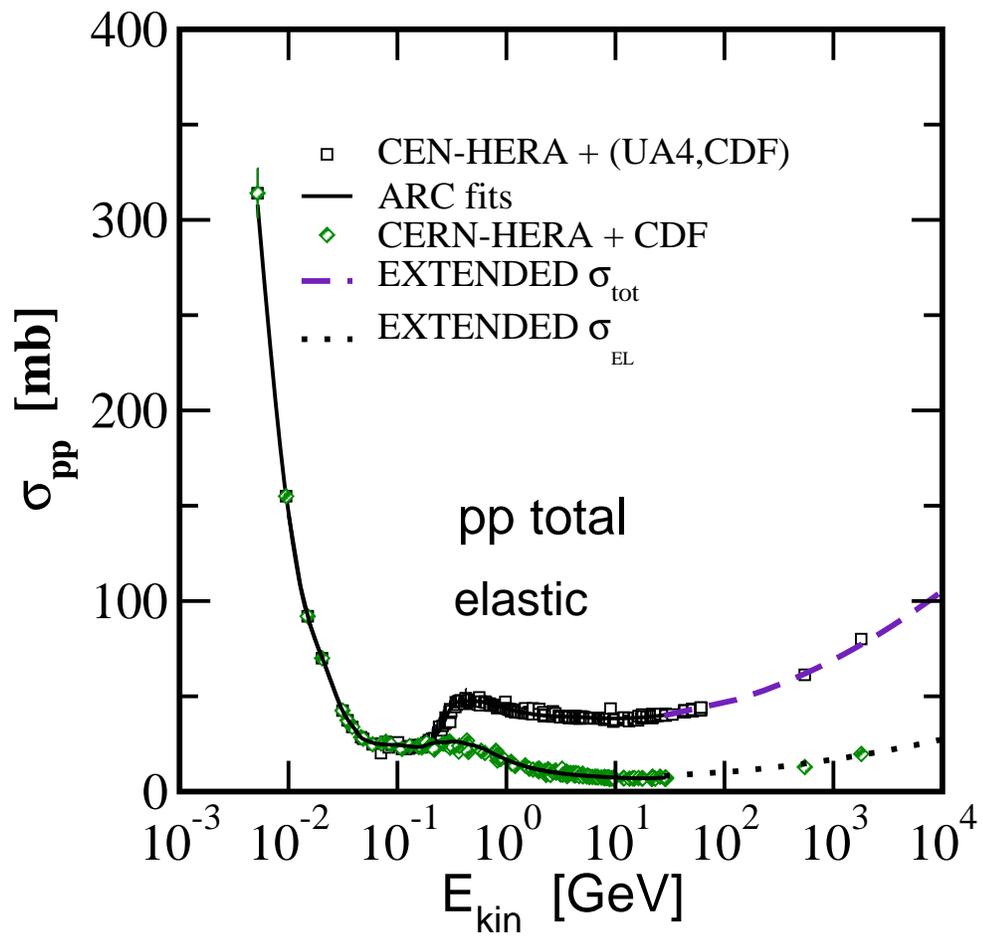}
\hfil}}
\caption[]{Total and  Elastic cross-sections: Comparison  of data
 compilations from  a variety of sources,  CERN-HERA, UA1,4,5 and
 CDF, with the fits used in LUCIFER.}
\label{fig:Fig.(2)}
\end{figure}
\clearpage 

\begin{figure}
\vbox{\hbox to\hsize{\hfil
\epsfxsize=6.1truein\epsffile[0 0 561 751]{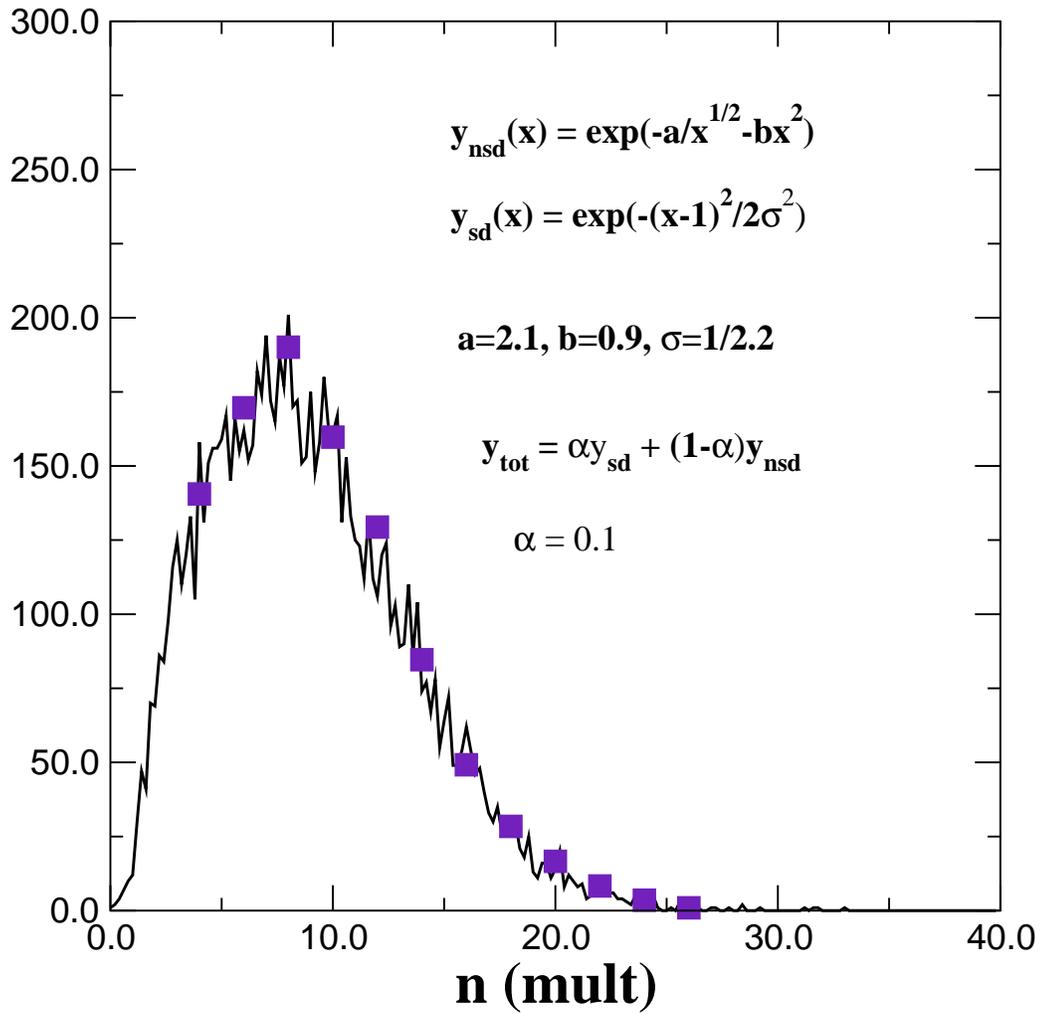}
\hfil}}
\caption[]{Multiplicity Distributions: Data for $s^{1/2}$ = 26 GeV is
 compared to a histogram using the KNO scaling outlined on this
 figure. The histogram is obtained from a reasonable number of pp
 events, and of course is itself normalized to a probability
 distribution.  Clearly the same KNO distribution is used at all
 energies with $x= n/\bar n$.}
\label{fig:Fig.(3)}
\end{figure}
\clearpage

\begin{figure}
\vbox{\hbox to\hsize{\hfil
\epsfxsize=6.1truein\epsffile[0 0 561 751]{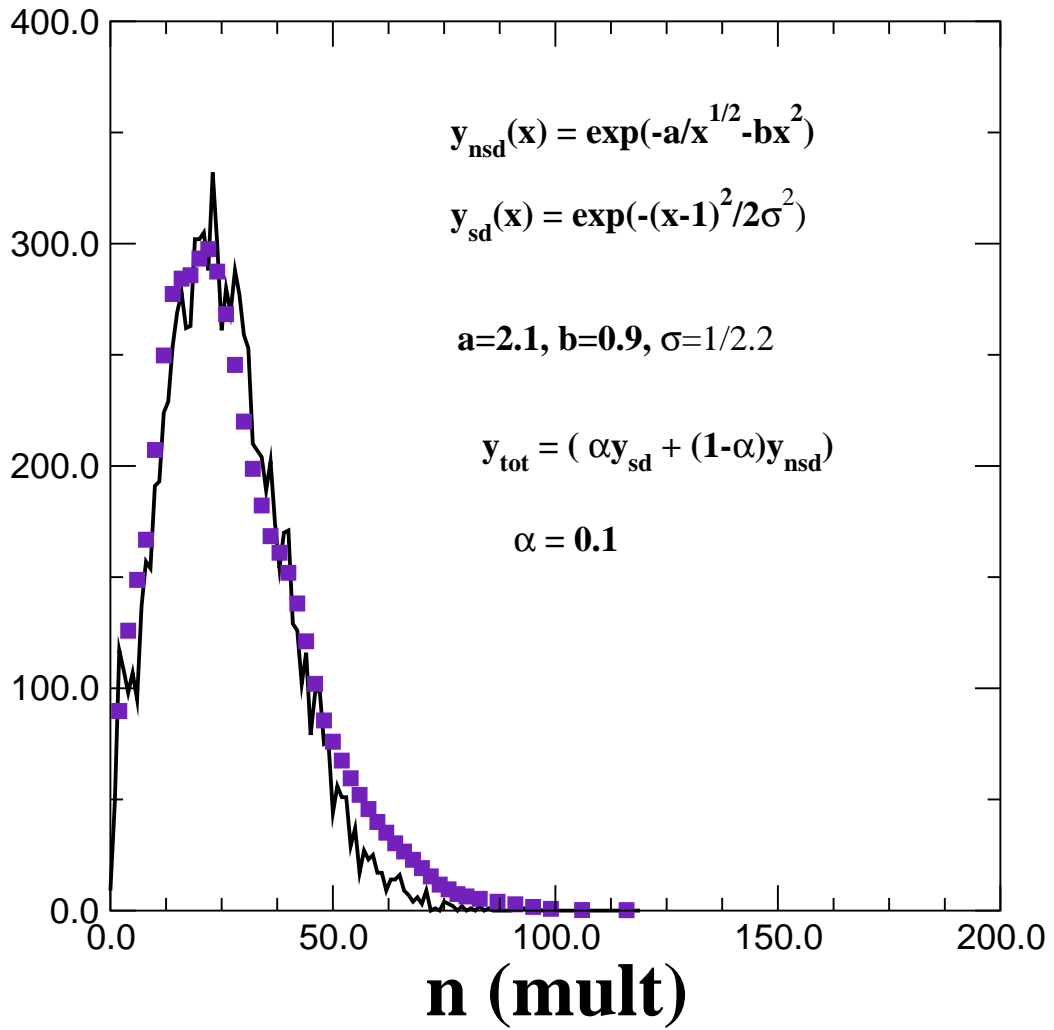}
\hfil}}
\caption[]{Multiplicity Distributions: Data for $s^{1/2}$ = 546 GeV is
 compared to a histogram using the KNO scaling outlined on this
 figure.  The strict KNO scaling seems to fail in the high
 multiplicity tail of the distribution shown here, but in a fashion
 which very little influences our A+A simulations.  Clearly the same
 KNO distribution is used at all energies.}
\label{fig:Fig.(4)}
\end{figure}
\clearpage

\begin{figure}
\vbox{\hbox to\hsize{\hfil
\epsfxsize=6.1truein\epsffile[0 0 561 751]{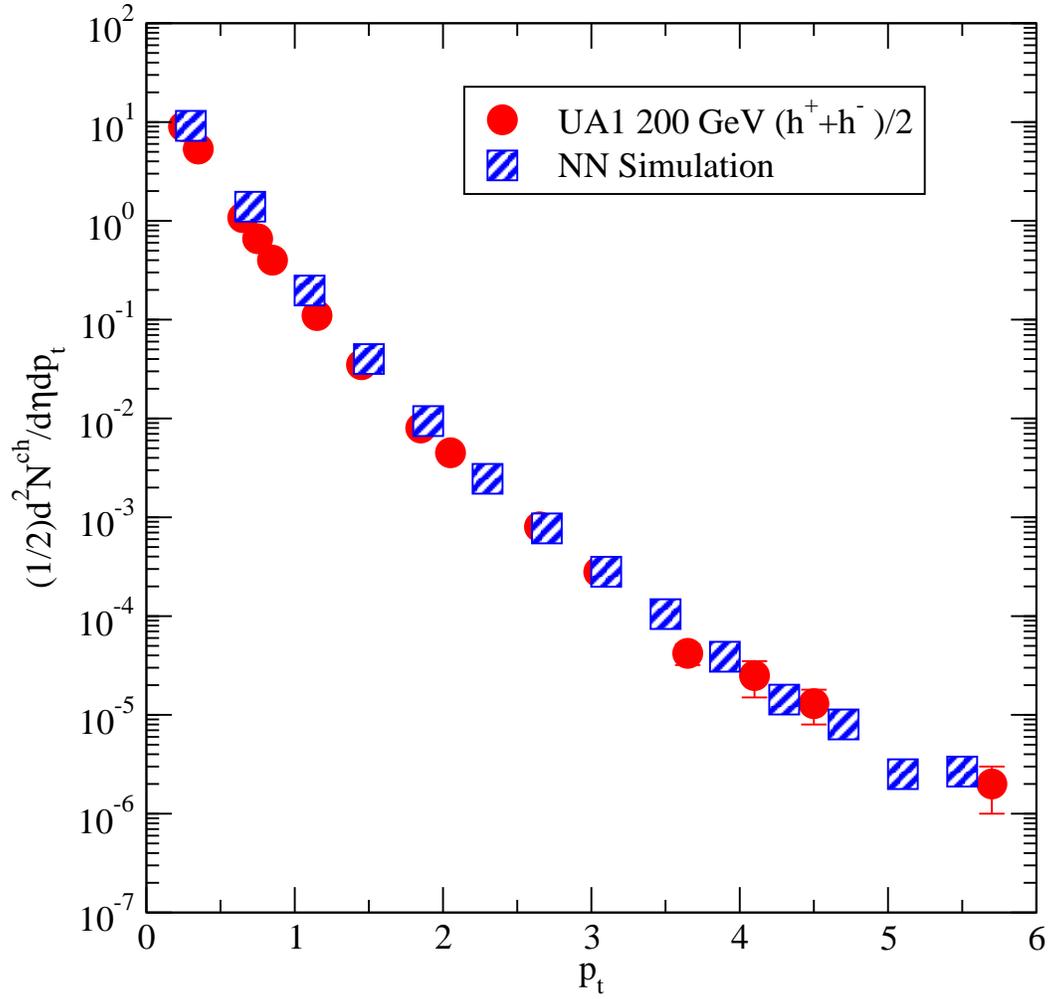}
\hfil}}
\caption[]{pp Pseudo-rapidity spectra: Comparison of UA1 minimum bias
200 GeV NSD data~\cite{ua1} with an appropriate LUCIFER
simulation. The latter is properly constrained by experiment and is an
input to the ensuing AA collisions; thus does not constitute a `set'
of free parameters.}
\label{fig:Fig.(5)}
\end{figure}
\clearpage

\begin{figure}
\vbox{\hbox to\hsize{\hfil
\epsfxsize=6.1truein\epsffile[0 0 561 751]{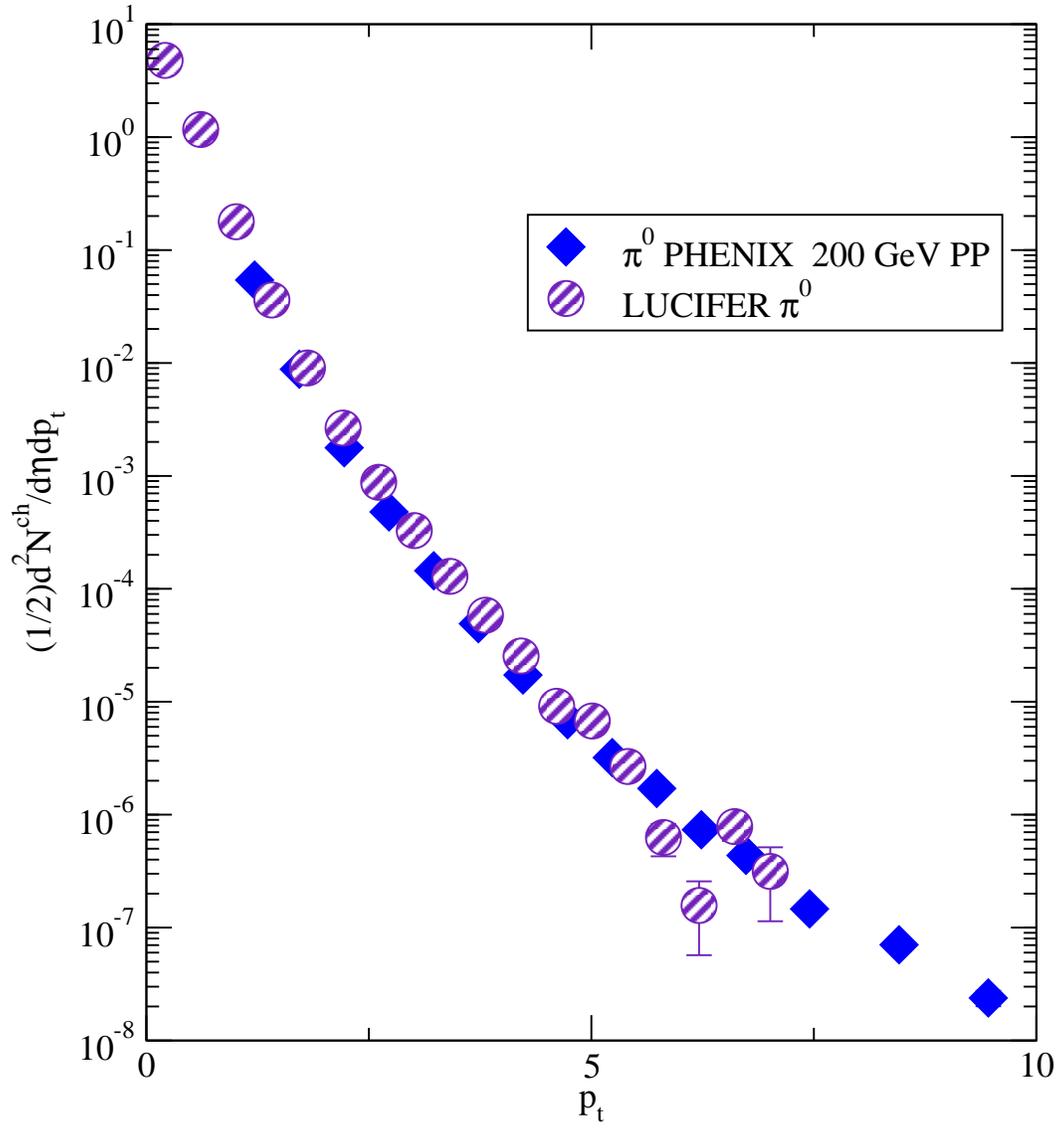}
\hfil}}
\caption[]{A  similar transverse  momentum  $\pi^0$ spectrum  from
PHENIX pp~\cite{phenixpp} vs simulation.}
\label{fig:Fig.(6)}
\end{figure}
\clearpage

\begin{figure}
\vbox{\hbox to\hsize{\hfil  
\epsfxsize=6.1truein\epsffile[0 0 561 751]{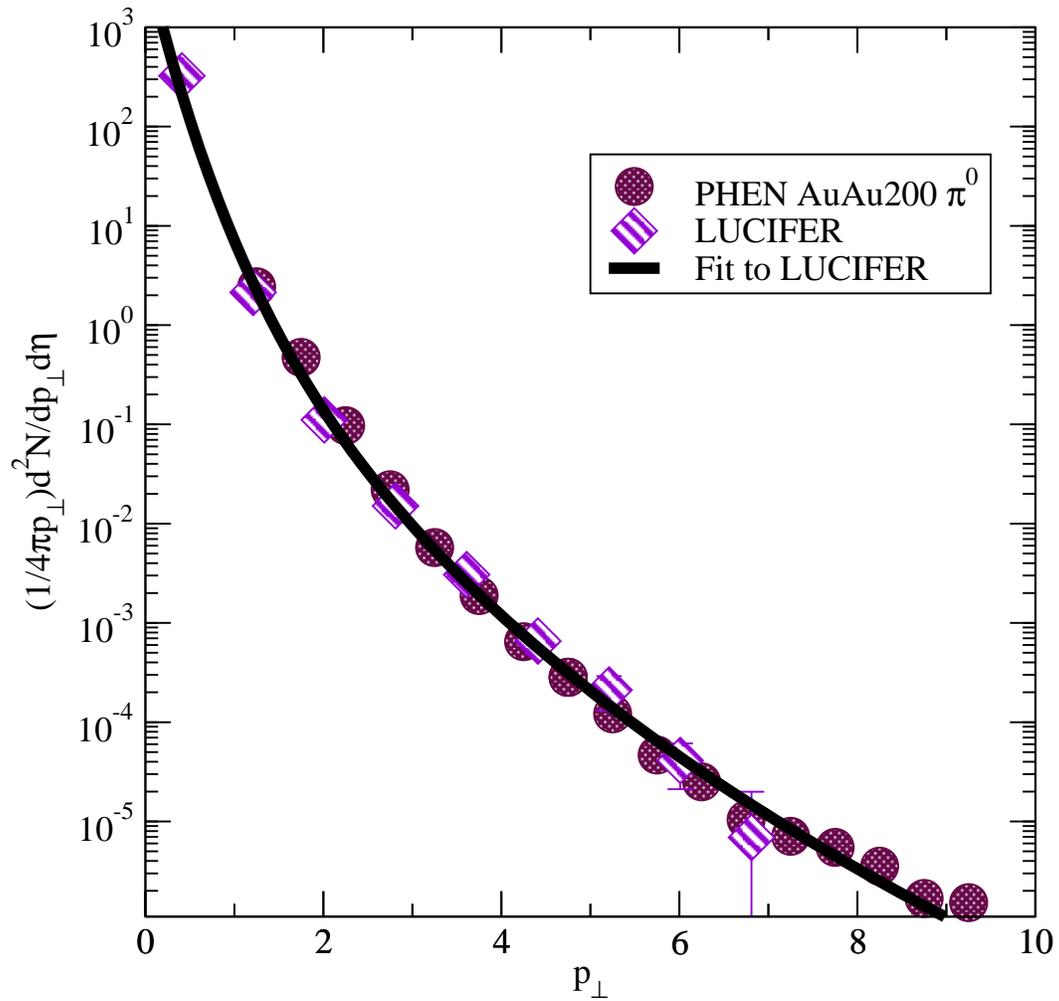} 
\hfil}}
\caption[]{Central PHENIX $\pi^0$ 200 GeV for Au+Au vs simulation.
Curves for different choices of the production time $\tau_p$ differ
very little, since in effect the cascade effectively begins somewhat
later, near $0.25-0.35$ fm/c and continues much longer to tens of
fm/c.  Centrality for PHENIX is here $0\%-10\%$, roughly for impact
parameters $b<4.25$ fm. in the simulation.}
\label{fig:Fig.(7)}
\end{figure}
\clearpage

\begin{figure}
\vbox{\hbox to\hsize{\hfil
\epsfxsize=6.1truein\epsffile[0 0 561 751]{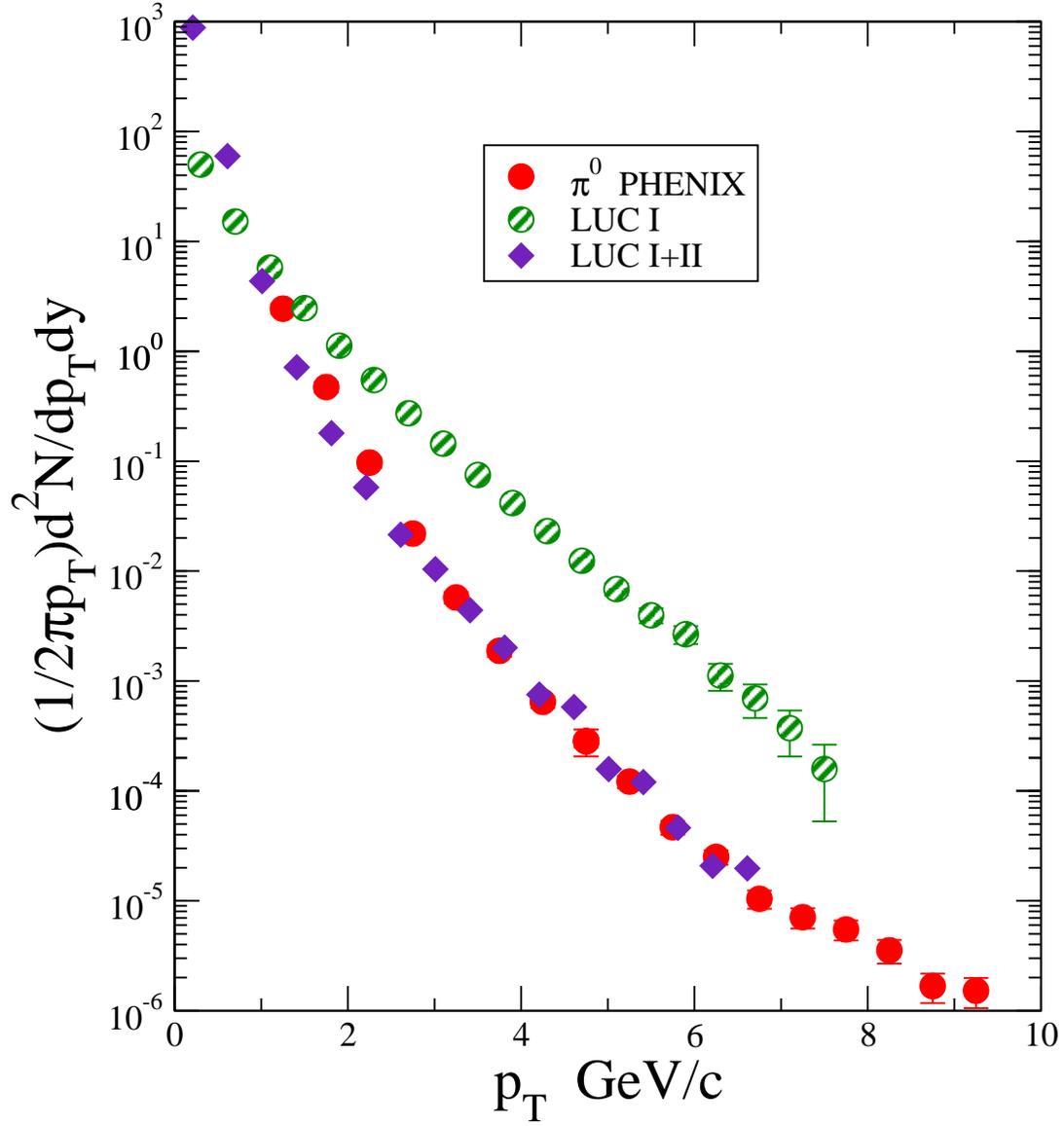}
\hfil}}
\caption[]{The $\pi^0$ transverse momenta yields for phase I, no final
cascade, vs those for the full phase I+II calculation.  Clearly there
is considerable suppression in the final cascade.  Recalling that the
experimentalists quote a `direct' suppression of $\sim$ 4--5 for the
ratio in Eqn.(1) at the highest $p_\perp$, there is at the end of I an
enhancement $\sim $3, {\it i.~e.}  still a Cronin effect in this first
phase.}
\label{fig:Fig.(8)}
\end{figure}
\clearpage

\begin{figure}
\vbox{\hbox to\hsize{\hfil
\epsfxsize=6.1truein\epsffile[0 0 561 751]{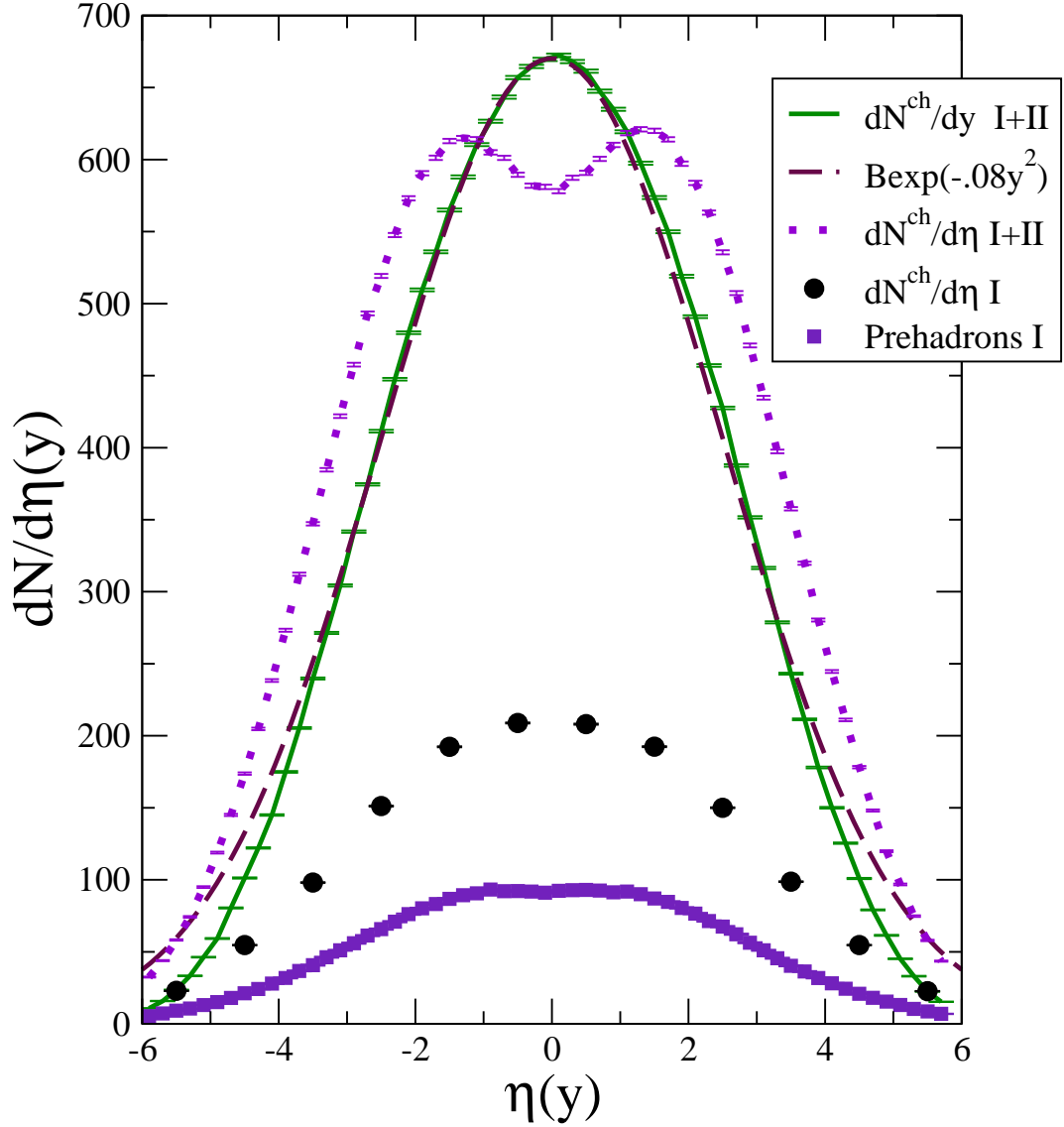}
\hfil}}
\caption[]{Pseudo-Rapidity and rapidity spectra for charged mesons and
pre-hadrons at various phases of the collision simulation.  The
Gaussian fit (dashed line) to the charged pion rapidity distribution
approximates preliminary BRAHMS $dN^{ch}/dy$~\cite{brahmsfwhmprelim}
results, certainly in its FWHM, and hence demonstrates the validity of
the simulation (solid line) for $dN^{ch}/dy$.  A calculation for
$dN^{ch}/d\eta$ (dotted line) is also presented.  For the charged
pseudo-rapidity distributions, successive retreats, first to phase I
of the simulation (solid circles) and then to only pre-hadrons in
phase I (solid rectangles), {\it i.~e.}  no decays, indicates both the
reduction in cascading participators and in the fraction of $E_\perp$
available at the earliest moments of the cascade.}
\label{fig:Fig.(9)}
\end{figure}
\clearpage

\begin{figure}
\vbox{\hbox to\hsize{\hfil
\epsfxsize=6.1truein\epsffile[0 0 561 751]{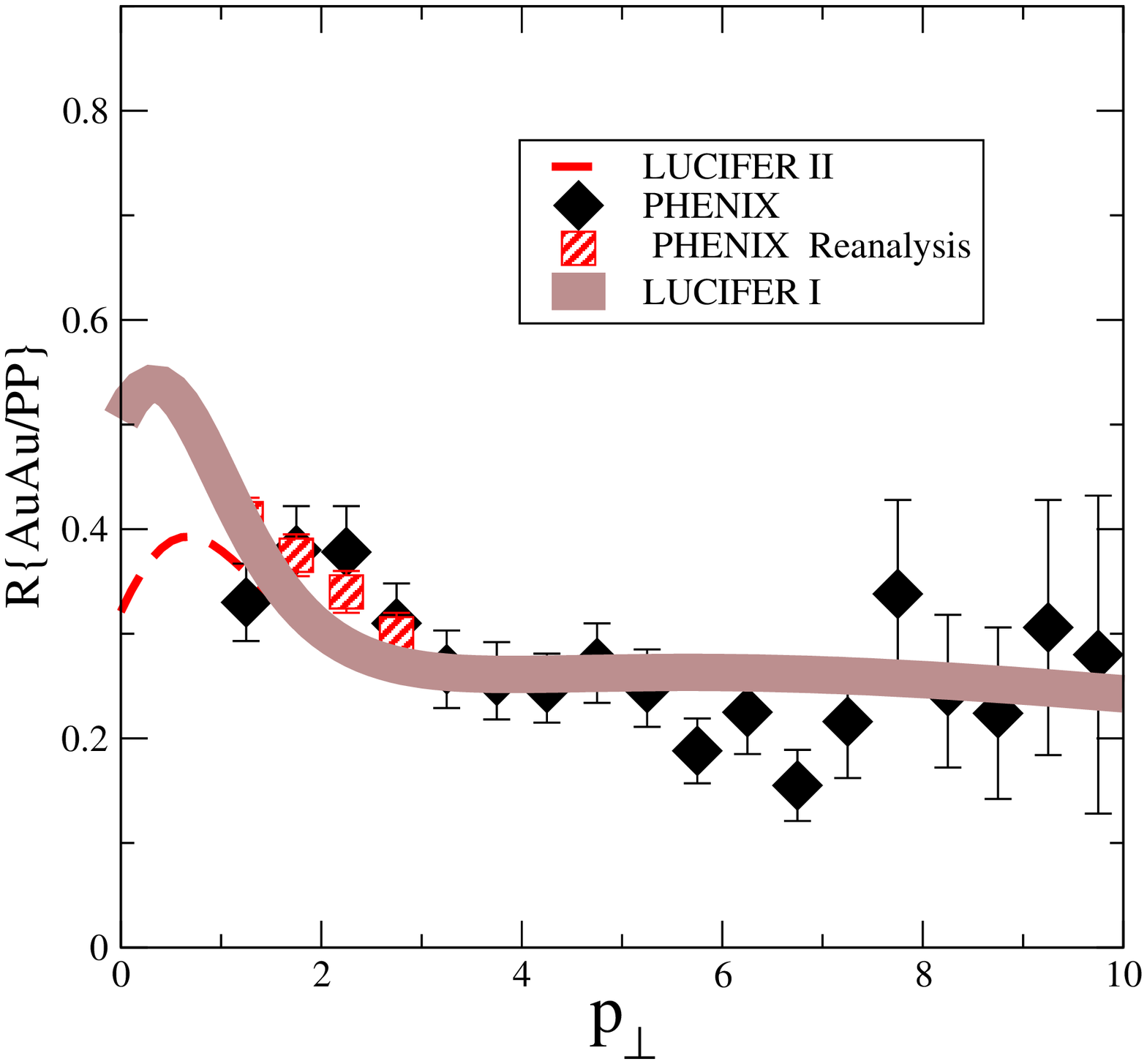}
\hfil}}
\caption[]{${R}_{AuAu/NN}$: Ratio of $\pi^0$ production in $Au+Au$
collisions to that in NN at ${s}^{1/2}$= 200 GeV for $0\%-10\%$
centrality. The binary collision number selected at this centrality is
N$_{coll}$=950, while the divisor for both PHENIX and LUCIFER curves
shown here are ones fit to the PHENIX pp data.  The resultant ratio is
especially sensitive to the PHENIX NN data. LUCIFER curves are shown
for two choices A and B differing not in the $Au+Au$ calculations but
only in the fitted PHENIX NN extrapolation to lowest $p_\perp$, where
of course there is no data.  The extra points for the PHENIX AuAu
ratio at the four lowest $p_\perp$ GeV/c are from more recent
analysis~\cite{shimamura}.  Thus the general shape and magnitude of
the simulation conforms well with this most recent data analysis, both
exhibiting a rise toward smaller $p_\perp$.  This feature is also
present in the very similar $0\%-5\%$ measurement~\cite{shimamura}.
The calculated suppression is somewhat excessive at low $p_\perp$.}
\label{fig:Fig.(10)}
\end{figure}
\clearpage

\begin{thebibliography}{99}

\bibitem{luc4brahms}

D.~E.~Kahana and  S.~H.~Kahana, nucl-th/0406074; Phys.~Rev.~C{\bf
72}, 024903, (2005).
 

\bibitem{brahms}

I.  Arsene  {\it et al},  BRAHMS Collaboration, Phys.~Rev.~Lett.,
{\bf   91}  072305,   (2003);   R.~Debbe,  BRAHMS   Collaboration
nucl-ex/0403052;   I.~Arsene   {\it    et   al.},   the   BRAHMS,
nucl-ex/0307003;   I.~Arsene    {\it   et   al.},    the   BRAHMS
Collaboration,   nucl-ex/0403050;   I.~Arsene   {\it   et   al.},
nucl-ex/0307003.


\bibitem{phobos}
 

B.~B.~Back {\it  et al}, PHOBOS  Collaboration, Phys.~Lett., {\bf
461}, 297,  (2004); B.~B.~Back {\it  et al.}; B.~B.~Back  {\it et
al.},  the  PHOBOS   Collaboration,  Phys.~Rev.~Lett.,  {\bf  91}
072302-1,(2003).

\bibitem{phenix}

S.~S~Adler {\it et al},  PHENIX Collaboration, Phys.~Rev. C. {\bf
69},  034910, (2004);  Phys.~Rev.  lett.   {\bf 91},  072301, 2003;
K.~Adcox {\it et al}, nucl-ex/0207009.

\bibitem{star}

C.~Adler {\it  et al}, STAR  Collaboration, Phys.~Rev.~Lett. {\bf
91}, 172302,  2003; Phys.~Rev.~Lett.  {\bf 89  }, 202301, (2002);
J.~Adams {\it et al}, Phys.~Rev.~Lett. {\bf 91}172302,2005.





\bibitem{lucifer1}

D.~E.~Kahana  and  S.~H.~Kahana,  {\it Proceedings,  RHIC  Summer
Study'96},  175-192,  BNL,  July  8-19,  1996;  D.~E.~Kahana  and
S.~H.~Kahana, Phys.~Rev. C{\bf 58}, 3574 (1998); Phys.~Rev. C{\bf
59},1651 (1999).

\bibitem{lucifer2}

D.~E.~Kahana, S.~H.~Kahana, Phys.~Rev.~C{\bf 63}, 031901(2001).

\bibitem{luc3}

Proc. International Conference on  the Physics of the Quark-Gluon
Plasma, Ecole Polytechnique,  Palaiseau, France, Sept. 4-7, 2001;
D.~E.~Kahana, S.~H.~Kahana, nucl-th/0208063.


\bibitem{zahed}
 
E.~V.~Shuryak and I.~Zahed, hep-ph/0307267; and hep-th/0308073.

\bibitem{lattice}

S.~Datta,     F.~Karsch,     P.~Petreczky    and     I.~Wetzorke,
hep-lat/0208012; hep-lat/0403017; hep-lat/0309012.




\bibitem{boris1}

B.~Z.~Kopeliovich, J.~Nemchik,  I.~Schmidt, Nucl.~Phys. A{\bf782}
224-233, 2007.



\bibitem{berger}

E.~Berger. Z.~Phys C{\bf4} 289,1984.

\bibitem{niedermayer}

F.~Niedermayer, Phys.~Rev. D{\bf34} 3494, 1986.

\bibitem{glauber}

R.~J.~Glauber, in {\it Lectures in Theoretical Physics, edited by
W.~E.~Brittin  et  al.},  Interscience,  New York,  1959,  Vol.I,
p.315.

\bibitem{molnar}

D.~Molnar  and   M.~Gyulassy,  nucl-th/0102031;  nucl-th/0104018;
D.~Molnar,  presentation at  workshop  on "Creation  and Flow  of
Baryons in Hadronic and  Nuclear Collisions", ECT, Trento, Italy,
May 3-7, 2004 (unpublished).

\bibitem{flow}

S.~S.~Adler   {\it    et   al.},   the    PHENIX   Collaboration,
Phys.~Rev.~Lett. {\bf 91}, 182301  (2003); C.~Adler {\it et al.},
the STAR Collaboration, Phys.~Rev.~Lett. {\bf 87}, 182301 (2001),
S.~Manly,  the  PHOBOS  Collaboration,  Proceedings of  the  20th
Winter  workshop on Nuclear  Dynamics, Trelawney  Beach, Jamaica,
March15-20, 2003.


\bibitem{rqmd}
H.~Stoecker  and  W.~Greiner, Phys.~Rep.  {\bf  137}, 277  (1986;
R.~Matiello,      A.~Jahns,     H.~Sorge      and     W.~Greiner,
Phys.~Rev.~Lettt.,{\bf 74}, 2180 (1995).

\bibitem{rqmd2}

H.~Sorge, Phys.~Rev. C{\bf 52}, 3291 (1995).

\bibitem{bass1}

S.~A.~Bass {\it et al.}, Nucl.~Phys. A{\bf 661}, 205 (1999).

\bibitem{frithjof}

B.~Andersson,   G.~Gustafson,   G.~Ingleman,  and   T.~Sjostrand,
Phys.~Rep  {\bf 97}, 31  (1983); B.~Andersson,  G.~Gustafson, and
B.~Nilsson-Almqvist,  Nucl.~Phys B{\bf 281}, 289 (1987).

\bibitem{capella}

A.~Capella and  J.~Tran Van,  Phys.~ Lett.~B{\bf 93},  146 (1980)
and Nucl.~Phys.~A{\bf 461}, 501c (1987); A.~Capella {\it et al.},
nucl-th/0405067 and hep-ph/0403081.

\bibitem{werner}

K.~Werner,  Z.~Phys. C{\bf  42},   85  (1989);  K.~  Werner,  J.~
Aichelin,  Phys.~Rev.~Lett.  {\bf   76}  (1996)  1027-1030;  H.~J.~
Drescher,  M.~Hladik,  S.~Ostapchenko,  K.~Werner, Proc.  of  the
``Workshop   on   Nuclear   Matter   in  Different   Phases   and
Transitions'', Les  Houches, France, March  31 - April  10, 1998;
K.~Werner  Invited lecture,  given at  the  Pan-American Advanced
Study Institute  "New States of Matter  in Hadronic Interactions"
Campos de Jordao, Brazil, January 7-18,2002, hep-ph/0206111.

\bibitem{ko}

B.~Zhang, C.~M.~Ko, B-A,~Li,  Z.~Lin, nucl-th/9904075; Z.~Lin and
C.~M.~Ko,  Phys.Rev.  C{\it  68}, 054904  (2003); Z.~Lin  {\it et
al.}  Nucl.Phys. A {\bf 698}, 375-378 (2002).

\bibitem{ranft}

J.~Ranft  and  S.~Ritter,  Z.~Phys.~C{\bf  27}, 413  (1985);
J.~Ranft   Nucl.~Phys.A{\bf 498},  111c (1989). 


\bibitem{boal} D.~Boal, {\it Proceedings of the RHIC Workshop I},
  (1985) and Phys.~Rev. C{\bf 33}, 2206 (1986).

\bibitem{eskola}

K.~J.~Eskola,  K.~ Kajantie  and J.~Lindfors,  Nucl.~Phys.  B{\bf
323}, 37 (1989).

\bibitem{wang}

X.~-N.~Wang and  M.~Gyulassy, Phys.~Rev. D{\bf  44}, 3501 (1991);
X.~-N.~Wang,  {nucl-th/000814}  and nucl-th/0405029;  X.~-N.~Wang
hep-ph/0405125.

\bibitem{wang2}

M.~Gyulassy and X.~N.~Wang, Comp.~Phys.~Comm. {\bf83}, 307
(1994), 

\bibitem{geiger}

K.~Geiger and B.~Mueller Nucl.~Phys.   B{\bf 369}, 600 (1992); K.
Geiger   Phys.~Rev.    D{\bf   46},   4965   and   4986   (1992);
K.~Geiger,{\it  Proceedings  of  Quark Matter'83},  Nucl.~Phys.~A
{\bf 418},  257c (1984);  K.~Geiger, Phys.~Rev.~D {\bf  51}, 2345
(1995).

\bibitem{bass2}

S.~A.~Bass,  B.~Mueller  and D.~K.~Srivastava,  Phys.~Rev.~Lett.B
{\it 551},277 (2003); S.~A. Bass {\it et al.}, Nucl.~Phys. A{\bf
661}, 205 (1999).

\bibitem{greiner}

K.~Gallmeister,  C.~Greiner  and  Z.~Xu, Phys.~Rev.~C{\it  67},
044905 (2003).

\bibitem{cassing}

W.~Cassing,  K.~Gallmeister, C.~Greiner Nucl.Phys.  A735, 277-299
(20004); J.~Geiss,  C.~Greiner,  E.~Bratkovskaya,  and  U.~Mosel,
Phys.~Lett. B{\bf 447}, 31 (1999).
 
\bibitem{wolschin}

G.~Wolschin,  Europhy.~Lett  {\bf  74},  (2006),  29-35;  Annelen
Phys.  {\bf  15},(2006),  369-378;  Phys.~Rev.  {\bf  69},(2004),
024906.


\bibitem{saturation}

L.~V.Gribov, E.~M.Levin and M.~G.~Ryskin, Phys.~Rep. {\bf 100}, 1
(1983); Nucl.~Phys.  B{\bf 188},  555  (1981); A.~H.~Mueller  and
J.~Qiu,  Nucl.~Phys.   B{\bf  268},427  (1986);   E.~M.Levin  and
M.~G.~Ryskin, Phys.~Rep. {\bf 189}, 267 (1990).

\bibitem{cgc1}

L.~McLerran  and  R.~Venugopalan,   Phys.~Rev.  D{\bf  49},  2223
(1994); Phys.~Rev.   D{\bf  59},   094002   (1999);  D.~Kharzeev,
E.~Levin and L.~McLerran, Phys.~Lett. B{\bf 561}, 93 (2003).

\bibitem{cgc2}

D.~Kharzeev,  E.~M.Levin and  L.~McLerran  hep-ph/0210332 (2002);
L.~McLerran  hep-ph/0402137 (2004);  D.~Kharzeev,  E.~M.Levin and
L.~McLerran, hep-ph/0403271 (2004).

   
\bibitem{goulianos}

K.~Goulianos, {\it Phys.~Rep.}{\bf 101}, 169 (1983).

K.~Goulianos, {\it Phys.~Rep.}{\bf 101}, 169 (1983).

\bibitem{COMPAS}

V.~Ezhela,  S.~Lugovsky,  N.Tkachenko,  and  Yu.~Kuyanov.   IHEP,
Protvino, Russia, August 2005, in The review of particle Physics,
W-M Yao et al. Phys.G33(2006).


\bibitem{CDF}        

Giorgio Chiarelli for the CDF Collaboration, {\it Proceedings of
the Topical Workshop on Proton-Antiproton Collider Physics}
October 18-22, 1993. Tsukuba, Japan; F.~Abe {\it et al.}
Phys.~Rev. D50,550-5561,1994.



\bibitem{kno}

Z.~Koba,H.~~B.~Nielsen, and P.~Olesen,Nucl.~Phys., B40 317, 1972.

\bibitem{intermittency}

UA5 Collaboration, Phys.~Lett. B{\bf 561}, 93 (2003).

\bibitem{ua5}

G.~Ekspong for the UA5 Collaboration, Nucl.~Phys.A{\bf 461}, 145c
(1987); G.~J.~~Alner for  the UA5 Collaboration, Nucl.~Phys.B{\bf
  291}, 445 (1987).

\bibitem{ua1}

C.~Albajar {\it et al.}, the UA1 Collaboration, Nucl.~Phys. B{\bf
335}, 261-287 (1990). 



\bibitem{fermilab}

Y.~Eisenberg {\it  et al.},  Nucl.~Phys. A{\bf 461},  145c (1987)
G.~J.~Alner {\it  et al.}, Nucl.~Phys.  B {\bf 291},  445 (1987);
F.~Abe et al., {Phys.~Rev.}{\bf D41} 2330, (1990).



\bibitem{phobos1}


B.~B.~Back  {\it et  al.}, the  PHOBOS  Collaboration, Phys.~Rev.
~Lett. {\bf 88}, 22302 (2002).



\bibitem{cronin}

J.~W.Cronin {\it et al.}, Phys.~Rev. D{\bf 91}, 3105 (1979).




 







\bibitem{gottfried}

K.~Gottfried,  Phys.~Rev.~Lett.    {\bf  32},  957   (1974);  and
Acta.~Phys.~Pol B{\bf 3}, 769 (1972).

\bibitem{phenixpp}

S.~S.~Adler     {\it     et     al},    PHENIX     Collaboration,
Phys.~Rev.~Lett. {\bf 91} 241803, (2003).



\bibitem{brahmsfwhmprelim}

I.~L.~Arsene     {\it    et     al},     BRAHMS    Collaboration,
nuc-ex/0403050,(2004).

\bibitem{phenix4}

K.~Adcock {\it et al}, PHENIX Collaboration, nucl-ex/0207009.

\bibitem{shimamura}

M.~Shimamura, for the PHENIX Collaboration, nucl-ex/0510023.






\end{thebibliography}
\end{document}